# Magic angle of Sr$_2$RuO$_4$: Optimizing correlation-driven superconductivity

Jonas B. Profe,[1,2,*] Luke C. Rhodes,[3,*] Matteo Dürrnagel,[4,5,*] Rebecca Bisset,[3] Carolina A. Marques,[6] Shun Chi,[7,8] Tilman Schwemmer,[4] Ronny Thomale,[4] Dante M. Kennes,[2,9] Chris A. Hooley,[10] and Peter Wahl[3,11]

[1]*Institute for Theoretical Physics, Goethe University Frankfurt, Max-von-Laue-Straße 1, D-60438 Frankfurt a.M., Germany*
[2]*Institut für Theorie der Statistischen Physik, RWTH Aachen University and JARA—Fundamentals of Future Information Technology, 52056 Aachen, Germany*
[3]*SUPA, School of Physics and Astronomy, University of St Andrews, North Haugh, St Andrews, KY16 9SS, United Kingdom*
[4]*Institute for Theoretical Physics and Astrophysics, University of Würzburg, D-97074 Würzburg, Germany*
[5]*Institute for Theoretical Physics, ETH Zürich, 8093 Zürich, Switzerland*
[6]*Physik-Institut, Universität Zürich, Winterthurerstrasse 190, CH-8057 Zürich, Switzerland*
[7]*Department of Physics and Astronomy, University of British Columbia, Vancouver BC, Canada V6T 1Z1*
[8]*Stewart Blusson Quantum Matter Institute, University of British Columbia, Vancouver BC, Canada V6T 1Z4*
[9]*Max Planck Institute for the Structure and Dynamics of Matter, Center for Free Electron Laser Science, 22761 Hamburg, Germany*
[10]*Max Planck Institute for the Physics of Complex Systems, Nöthnitzer Straße 38, 01187 Dresden, Germany*
[11]*Physikalisches Institut, Universität Bonn, Nussallee 12, 53115 Bonn, Germany*

Understanding of unconventional superconductivity is crucial for engineering materials with specific order parameters or elevated superconducting transition temperatures. However, for many materials, the pairing mechanism and symmetry of the order parameter remain unclear: reliable and efficient methods of predicting the order parameter and its response to tuning parameters are lacking. Here, we investigate the response of superconductivity in Sr$_2$RuO$_4$ to structural distortions via the random phase approximation (RPA) and functional renormalization group (FRG), starting from realistic models of the electronic structure. Our results suggest that RPA misses the interplay of competing fluctuation channels. FRG reproduces key experimental findings. We predict a magic octahedral rotation angle, maximizing the superconducting $T_c$ and a dominant $d_{x^2-y^2}$ pairing symmetry. To enable experimental verification, we provide calculations of the phase-referenced Bogoliubov Quasiparticle Interference imaging. Our work demonstrates a designer approach to tuning unconventional superconductivity with relevance and applicability for a wide range of quantum materials.

DOI: 10.1103/PhysRevResearch.6.043057

## I. INTRODUCTION

Superconducting materials have the potential to revolutionize technology and everyday life [1,2] if we can engineer materials with specific order parameters or higher superconducting transition temperatures ($T_c$). To enable this, however, we first require a detailed understanding of the origin of a material's superconducting mechanism and order parameter, which for many materials is still lacking.

Strontium ruthenate, Sr$_2$RuO$_4$, is a case in point: despite over 30 years of research [3], there is still no consensus on the symmetry of its superconducting order parameter. This material was a particular focal point of research into unconventional superconductivity due to early suggestions of a spin-triplet superconducting order parameter [4–7]. However, a key piece of evidence in support of this claim, the absence of a Knight shift on entering the superconducting state [5], was later shown to be a consequence of experimental artefacts [8,9]. Nevertheless, there is still a significant body of experimental evidence [10] that supports an unconventional pairing symmetry.

Key experimental techniques that could resolve the superconducting pairing symmetry have failed to provide resolution. Via quasiparticle interference imaging, scanning tunneling microscopy (STM) has provided important insights into the pairing symmetries of the cuprate and iron-based superconductors [11,12], but appears blind to the superconductivity in Sr$_2$RuO$_4$ [13,14] and has failed to provide conclusive evidence for the symmetry [15,16]. The clean surface reconstructs inducing a rotation of the RuO$_6$ octahedra [17,18]. Evidence for superconducting gaps from STM has only been found on disordered or dirty surfaces [19–21] or in small nanometer-sized holes [22].

The theoretical description of superconductivity driven by electronic correlations, often hypothesized as the candidate pairing mechanism in Sr$_2$RuO$_4$, is not straightforward [23]. The correlations have to be treated with quantum many-body techniques, such as dynamical mean-field theory [24],

---

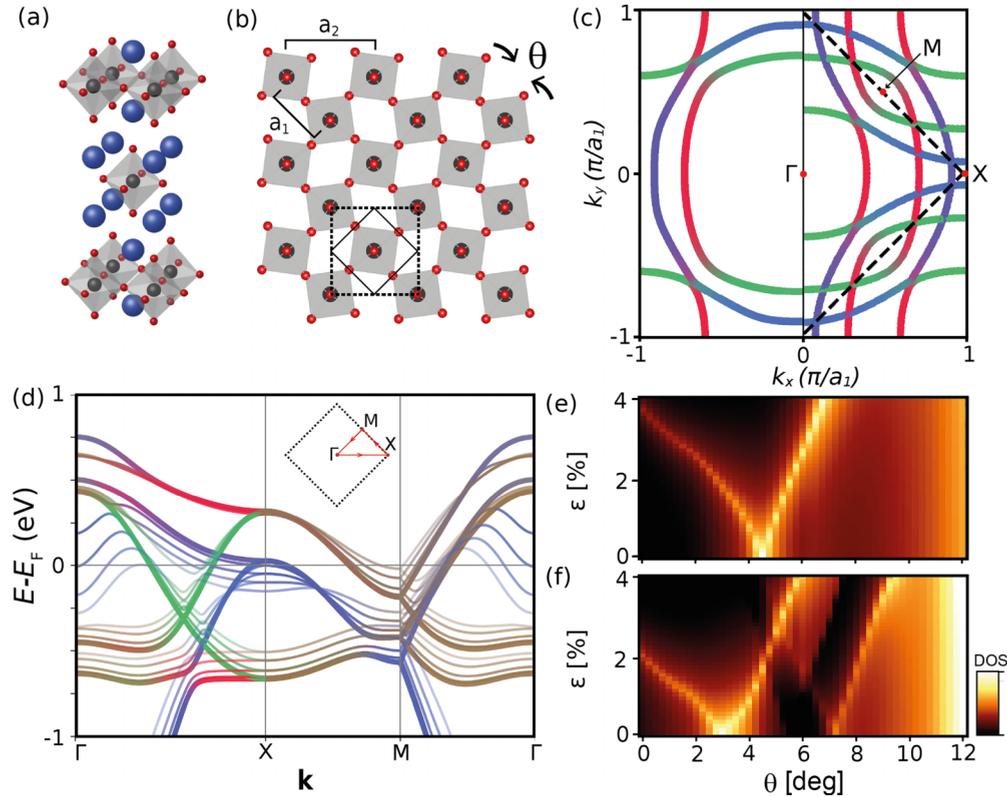

FIG. 1. Tuning electronic structure of $Sr_2RuO_4$ by octahedral rotation and uniaxial strain. (a) Crystal structure of bulk $Sr_2RuO_4$. (b) Top-down view of a single $RuO_4$ plane showing the effect of a finite in-plane rotation $\theta$ of the oxygen octahedra. The solid box shows a 1-Ru atom unit cell and the dashed box a 2-Ru atom unit cell required when finite octahedral rotation is present. $a_1$ and $a_2$ refer to the lattice constants of the 1 and 2 Ru atom unit cell. (c) Fermi surface of $Sr_2RuO_4$ without octahedral rotation using the 1-Ru atom unit cell (left), and the 2-Ru atom unit cell (right). The orbital character has been encoded via red, green, and blue for the $d_{xz}$, $d_{yz}$ and $d_{xy}$ orbitals, respectively. (d) Bandstructure along the irreducible path sketched in (c) as a function of octahedral rotation angle $\theta$. The thick line is the electronic structure for $\theta = 0°$, and the thin lines represent the dispersions with nonzero $\theta$, incremented in steps of $3°$. (e), (f) Total density of states (DOS) at the Fermi level $\rho(E_F)$ as a function of octahedral rotation $\theta$ and uniaxial strain $\epsilon$ along the nearest neighbor Ru-Ru direction, (e) without SOC, and (f) with SOC.

Quantum Monte Carlo [25], or, whenever applicable, diagrammatic methods [26–28], but the complexity of these techniques results in numerically challenging tasks for realistic multi-orbital systems with spin-orbit coupling such as $Sr_2RuO_4$.

In this work, we study the influence of small structural distortions of the crystal lattice of $Sr_2RuO_4$ on the expected superconducting order parameter and $T_c$, motivated by the sensitivity the superconductivity exhibits to uniaxial strain. Strain results in an increase of $T_c$ by up to a factor of two [29–31] driven via a Lifshitz transition when a Van-Hove singularity (VHs) crosses the Fermi energy $E_F$ [32]. Similar sensitivity might be expected with biaxial strain [33] and as a function of octahedral rotations as, e.g., in $Ca_{2-x}Sr_xRuO_4$ [34].

Our study is based on the assumption that the superconductivity in $Sr_2RuO_4$ is due to electronic correlation effects. We therefore use two numerical techniques well adapted to this case: the random phase approximation (RPA) [27,35], which has already been extensively used to study the superconductivity in $Sr_2RuO_4$ [36–42], and truncated unity functional renormalization group theory (TUFRG) [43–47].

The remainder of this article is organized as follows. In Sec. II we introduce the normal-state electronic structure for experimentally accessible rotations of the $RuO_6$ octahedra and uniaxial strain. In Secs. III and IV, we employ the RPA and TUFRG techniques, respectively, to investigate how structural distortions affect the pairing symmetry and transition temperature. Finally, in Sec. V, we provide specific tests for how the symmetry of the order parameter can be experimentally verified from simulations of Bogoliubov quasiparticle interference (BQPI). While the results in this paper are specific to $Sr_2RuO_4$, our work also establishes a computational framework for simulating electronic properties that can be applied to a wide range of materials.

## II. STRUCTURAL CONTROL OF ELECTRONIC STRUCTURE

In this section, we introduce the tight-binding models for the single-particle electronic structure used as a basis for the calculations of the superconducting instabilities of $Sr_2RuO_4$. The crystal structure of $Sr_2RuO_4$ is shown in Fig. 1(a). Given the quasi-two-dimensional nature of the electronic structure of $Sr_2RuO_4$, which is characterized by a 1000 times higher resistivity along the crystallographic $c$ direction than in the $a - b$-plane [48] and a negligible $k_z$-dispersion of the electronic bands in the vicinity of $E_F$ [49,50], we model the electronic structure by considering a single free-standing layer of Sr$_2$RuO$_4$.

### A. Influence of octahedral rotations

It is known that Sr$_2$RuO$_4$ exhibits a soft phonon mode associated with in-plane octahedral rotations [51] that freezes out at the surface of the material to produce a $\sqrt{2} \times \sqrt{2}$ reconstruction [18], as shown in Fig. 1(b). Apart from bands shifting due to the octahedral rotation, it results in a larger unit cell and folded Brillouin zone. To model the changes in the electronic structure associated with such rotation, we have performed Density Functional Theory (DFT) calculations for octahedral rotations between 0° and 12° in steps of 1°, while keeping the lattice constant to the equilibrium value at 0° [52]. This captures the behavior expected for a surface layer. These DFT calculations have been projected onto tight-binding models [53] retaining the $t_{2g}$ bands that cross $E_F$. Models for noninteger angles have been obtained by spline interpolation across the models. Spin-orbit coupling (SOC) has important consequences for the band structure of Sr$_2$RuO$_4$ [54,55] and is introduced at the level of the tight-binding model by adding a term $H_{\text{SOC}} = \lambda \mathbf{L} \cdot \mathbf{S}$, here with $\lambda = 175$ meV, slightly larger than what relativistic DFT calculations would suggest in accordance with experimental findings [56].

From the calculations, we can assess the consequences of the octahedral rotations for the electronic structure: (1) several bands are back-folded into the smaller Brillouin zone, as shown in Fig. 1(c), and (2) the hoppings of the in-plane $d_{xy}$-orbital are modified, inducing a sequence of Lifshitz transitions. The first Lifshitz transition is due to the VHs at the X point moving across $E_F$. It is followed by a SOC induced gap at about two thirds along $\Gamma - X$ that opens with increased octahedral rotation and results in a pair of Lifshitz transitions when the gap edges cross $E_F$. Finally, an electron pocket forms around the $\Gamma$ point, as shown in Fig. 1(d), due to the octahedral rotation generating finite overlap of the $d_{xy}$ and $d_{x^2-y^2}$ bands [57]. Experimental data, from ARPES and STM measurements, matches the qualitative features of the electronic structure shown here [52,55,56].

### B. Influence of uniaxial strain

In experiments, the largest impact of uniaxial strain is observed for strain along the [1 0 0] direction of the tetragonal unit cell [29,30,32]. Uniaxial strain can result in distortion of the RuO$_6$ octahedra, as well as in a change of rotation angle, with experiments suggesting that the dominant structural change is distortion [55]. Therefore, to introduce uniaxial strain, $\epsilon$, into our tight-binding models in a way that is consistent with experiments, we modify the nearest neighbor hopping parameters of the models by multiplying the hopping terms in the direction of the strain by $1 + \epsilon$, and divide them by $1 + \epsilon$ in the orthogonal direction. Comparison with experimental data of how strain affects the bulk and surface electronic structure allows us to express $\epsilon$ in terms of the strain $\epsilon^{\text{exp}} = (\epsilon_{xx} - \epsilon_{yy})$ applied in experiments (see, e.g., [32,55]). We obtain $\epsilon \approx 3\epsilon^{\text{exp}}$ in relation to the usual definition of strain $\epsilon^{\text{exp}}$, as used in experiments or when introducing strain directly into DFT calculations [58]. The most obvious effect of the uniaxial strain is to introduce a splitting of the $d_{xy}$ VHs that is located close to $E_F$ at the X point.

### C. Impact of octahedral rotation and uniaxial strain on the density of states

The presence of VHs's or hybridization gaps at $E_F$ has dramatic consequences for the density of states (DOS) $\rho(E)$, and can already serve as an indicator for the changes in the superconducting transition temperature with octahedral rotations and strain. In Figs. 1(e) and 1(f), we plot the DOS at the Fermi level, $\rho(E_F)$, as a function of uniaxial strain $\epsilon$ and octahedral rotation $\theta$. Without spin-orbit coupling [Fig. 1(e)], the VHs crosses the Fermi level at $\theta = 5°$, and splits under finite uniaxial strain. The inclusion of SOC [Fig. 1(f)] induces multiple additional features and modifications. Firstly, $\rho(E_F)$ reaches its maximum at a smaller rotation angle $\theta \sim 3°$ and $\epsilon = 0\%$. Secondly, the uniaxial strain required to reach the Lifshitz transition for $\theta = 0°$ is significantly reduced to $\epsilon = 2\%$. And thirdly, for angles between $5° \leqslant \theta \leqslant 7°$, $\rho(E_F)$ quickly decreases due to the SOC induced hybridization gap. At an angle of 6°, an additional electron pocket appears at the $\Gamma$ point that is also detected in ARPES measurements at the reconstructed surface [55], followed by another VHs at 12°. If the superconducting transition temperature was only dependent on the DOS at the Fermi level, we would therefore expect a maximum $T_c$ at $\theta = 12°$.

## III. SPIN FLUCTUATIONS AND LEADING PAIRING INSTABILITIES WITHIN THE RPA

As a result of the partial overscreening of the radially symmetric Coulomb repulsion by fluctuations of the electronic background in some angular momentum channel [59], superconductivity can emerge even in the absence of an attractive pairing interaction provided by an external degree of freedom, such as phonons or magnetic fluctuations from local moments. A prominent example is the nearest-neighbor attraction between electrons driven by spin fluctuations [60]. We will analyze the origin of an attractive Cooper pair interaction in the following using the random phase approximation (RPA) in the implementation presented in Ref. [61]. In Sr$_2$RuO$_4$, RPA has been extensively applied to identify the SC pairing symmetry [36,37,39,41,42], since the material features strong spin fluctuations due to a parametric vicinity to an antiferromagnetic Mott state [62]. To truncate the infinite hierarchy of higher order screening processes to the bare two particle interaction, the RPA assumes random phases of virtual excitations. This implies that in the thermodynamic limit all processes lacking phase coherence average out. As a direct consequence, only collective screening processes contribute to the effective interaction close to $E_F$ [63,64].

### A. Bare susceptibility profile

Away from $E_F$, the relevant screening processes are provided by virtual particle-hole (ph) fluctuations such as spin and charge density oscillations. In the static limit, their

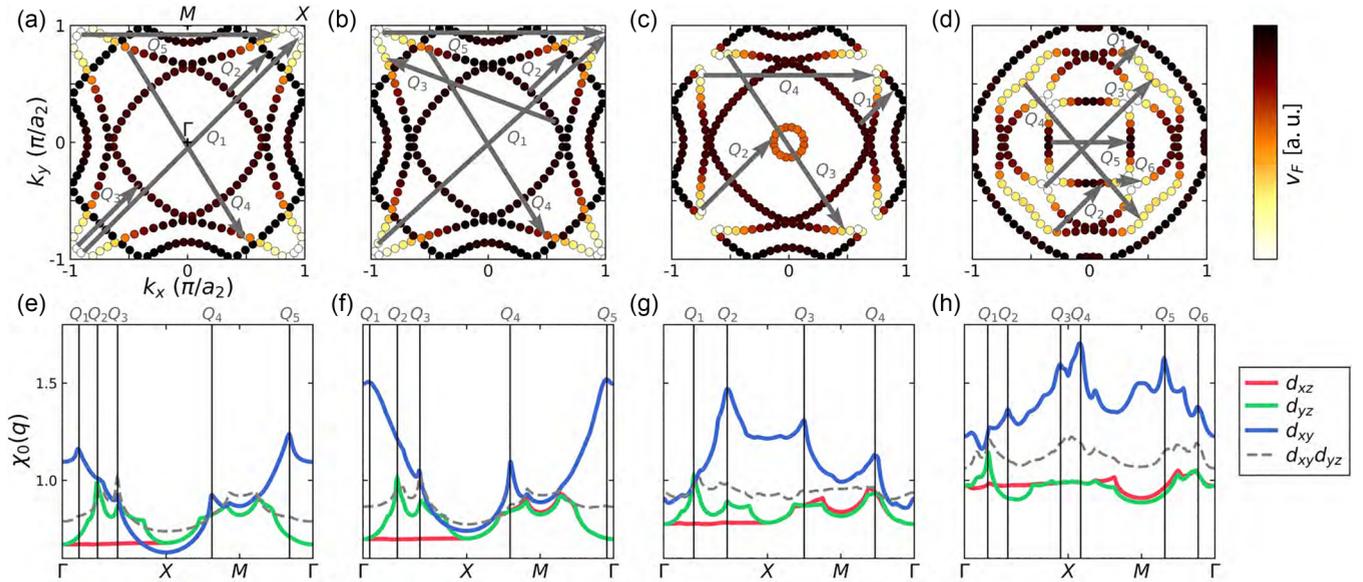

FIG. 2. Nesting in the band structure of $Sr_2RuO_4$ for different octahedral rotations. (a)–(d) FS topologies of $Sr_2RuO_4$ at different octahedral rotation angles, with (a) $\theta = 0°$, (b) $3°$, (c) $7°$, and (d) $12°$. The color bar encodes the Fermi velocity $v_F$ as a proxy for the inverse DOS, allowing to identify parts of the Fermi surface which contribute most to the nesting and hence to the bare susceptibility spectrum $\chi^0$. (e)–(h) The dominant orbital contributions to the spin susceptibility $\chi^0_{o_1 o_1 o_3 o_3}(\mathbf{q})$ are shown along the high symmetry path of the one-atom unit-cell.

strengths are given by the bare ph susceptibility

$$\chi^0_{o_1 o_2 o_3 o_4}(\mathbf{q}) = -\int_{BZ} \frac{d\mathbf{k}}{V_{BZ}} \frac{f(\beta\varepsilon_n(\mathbf{k}+\mathbf{q})) - f(\beta\varepsilon_m(\mathbf{k}))}{\varepsilon_n(\mathbf{k}+\mathbf{q}) - \varepsilon_m(\mathbf{k})} \times M^{nm}_{\{o_i\}}(\mathbf{k},\mathbf{q}), \quad (1)$$

that depends only on the single particle energies $\varepsilon_n(\mathbf{k})$ and orbital-to-band transformations $M^{nm}_{\{o_i\}}(\mathbf{k},\mathbf{q})$. The momentum space integral is normalized by the volume $V_{BZ}$ of the Brillouin zone (BZ). The Fermi distribution $f(\beta\epsilon)$ is evaluated at an inverse temperature $\beta$. Due to a vanishing denominator for $\varepsilon_n(\mathbf{k}+\mathbf{q}) \to \varepsilon_m(\mathbf{k})$, the fluctuation spectrum is highly susceptible to the precise shape of the Fermi surface and in particular nesting features thereof.

Inspecting the evolution of the Fermi surface (FS) with octahedral rotation $\theta$ and for zero strain, we can identify four distinct FS topologies shown in Figs. 2(a)–2(d). Disentangling the different contributions to the ph susceptibility allows us to pinpoint the effect of $\theta$ on the electronic fluctuations present in the system. Due to the enhanced DOS for the $d_{xy}$-derived bands in the vicinity of the VHs, the corresponding ph fluctuations dominate the susceptibility spectrum for all angles. However, the dominant susceptibility vectors $Q_i$ are substantially changing across the Lifshitz transitions as they open new and close existing ph scattering channels on the FS (cf. Fig. 2). This can be attributed to the sensitivity of $\chi^0$ to details and the geometry of the Fermi surface, and obscures, e.g., the nesting feature of the VHs at $X$ despite its vicinity to the Fermi level. The octahedral rotation hardly affects the quasi one dimensional $d_{xz}/d_{yz}$-bands apart from the hybridisation with the $d_{xy}$ orbitals. Hence, the considerable nesting of these bands persists up to $\theta = 12°$.

### B. Spin fluctuation mediated pairing

Moving from the single particle excitation spectrum provided by $\chi^0$ to the full many-body analysis, collective screening effects can be evaluated within the RPA in an analytic way up to infinite order via $\chi^{RPA} = \chi^0/(1 - \Gamma\chi^0)$. Throughout this work, we assume the two particle interaction $\Gamma$ to be independent of the octahedral rotation or strain. Following Ref. [65], we model the onsite interaction for the $t_{2g}$ manifold by a local Kanamori vertex with intra- and interorbital repulsions $U$, $V$ and interorbital Hund's coupling and pair hopping $J$ subject to the universal ratio $U = V + 2J$. We fix $J = 0.14U$. Further details on the bare interactions can be found in Sec. S3A of Supplemental Material [66]. To evaluate the superconducting pairing tendencies mediated by the ph fluctuations, the RPA susceptibility $\chi^{RPA}$ is treated as a screening background for the effective Cooper pair interaction on the FS and the superconducting gap is obtained within a mean field theory at the Fermi level [37,60,61]. The leading eigenvalue $\lambda$ of the linearised gap equation can be taken as a proxy for the expected superconducting transition temperature since $T_c \propto We^{-1/\lambda}$.

Since all higher order screening processes contributing to the effective Cooper pair vertex carry the symmetry of the initial Hamiltonian, the obtained superconducting order parameters can be classified by means of the irreducible representations (irreps) of the lattice point group $C_{4v}$, that can be expanded in a basis set of associated lattice harmonics [67,68]. Projected onto the FS, the superconducting order parameter retains dependence on pseudospin and Fermi momentum exclusively. This allows for a classification of the gap function into order parameters that are odd under spatial inversion and pseudospin triplet (two-dimensional $E$ irrep) and those that are even under spatial inversion and pseudospin singlet ($A_1$, $A_2$, $B_1$, $B_2$ irrep). The decomposition of the order parameter is detailed in Sec. S3B of Supplemental Material [66].

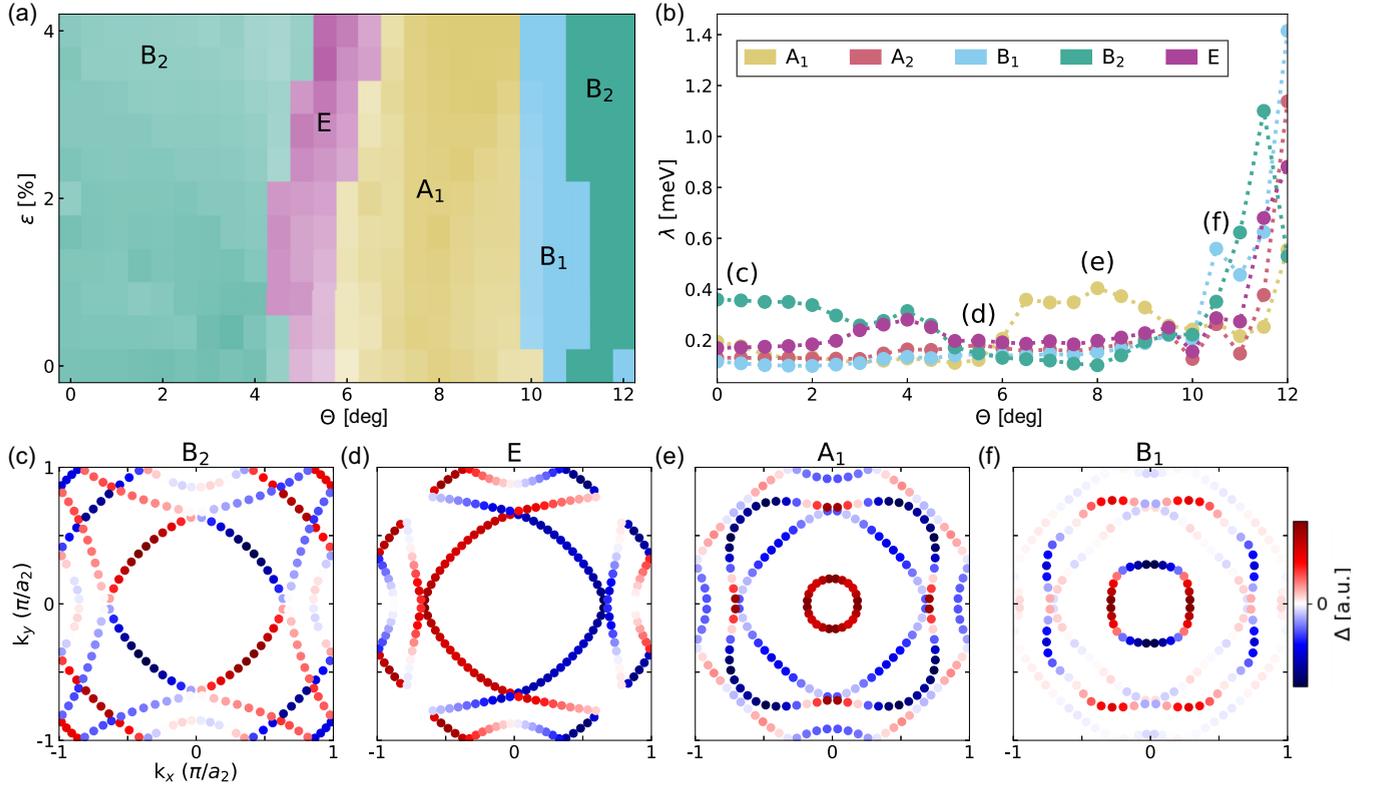

FIG. 3. Fermiology and superconducting pairing in Sr$_2$RuO$_4$ from RPA analysis. (a) The leading superconducting order is rapidly changing as a function of octahedral rotation $\theta$ and uniaxial strain $\epsilon$ displaying all irreps of the $C_{4v}$ point group in the phase diagram. Darker color indicates a larger pairing strength. (b) Line cut along $\epsilon = 0$ displaying no remarkable drop of $T_C$ for the experimentally observed surface rotation of $\theta \sim 8°$. The RPA results do not reproduce the experimentally observed absence of SC in the surface layer. (c)–(f) Exemplary gap functions for various angles as indicated in (b) show contributions to the SC order also from the circular Fermi pocket arising after the Lifshitz transition at $\theta = 5°$.

In the structural parameter space of rotation $\theta$ and strain $\epsilon$, the leading superconducting order obtained in RPA displays a variety of symmetries, compare Figs. 3(a) and 3(b), including pseudospin triplet pairing in the helical channel. The variety of orders stabilized here can be traced back to the competing effects of nesting from different parts of the Fermi surface and is therefore a consequence of changes of the FS topology with $\theta$ and $\epsilon$. The two most prominent effects are observed at the Lifshitz transitions as a function of $\theta$. When the $d_{xy}$-VHs comes close to the Fermi level, the perfect nesting of linear parts of the $d_{xy}$-derived Fermi surfaces promotes order parameters that are odd under inversion and hence pseudospin triplet in nature ($E$), different from the $d$-wave symmetry found at low angles $\theta$ [Figs. 3(c)–3(f)]. Directly after the emergence of the additional Fermi surface pocket around $\Gamma$, the associated DOS and its nesting to points close to the VHs promotes extended $s$-wave SC to efficiently gap out the circular pocket [Fig. 3(e)].

Remarkably, the observed superconducting pairing strength does not follow the DOS dependence of $\theta$ [compare Fig. 3(b) with Fig. 1(f)]. In particular, for angles smaller than 11°, the largest pairing is found at angles $\theta \sim 6 \ldots 9°$, in the range of octahedral rotations realized at the surface of Sr$_2$RuO$_4$ and where experimentally a suppression of $T_c$ is found [13,14,22]. This general trend also persists in the absence of SOC (see also Sec. S3C of Supplemental Material [66]).

## IV. SUPERCONDUCTING ORDER PARAMETER IN FRG

The seeming contradiction between the RPA results for the superconductivity for an electronic structure as found in the surface layer of Sr$_2$RuO$_4$ and the experimentally observed suppression of $T_c$ suggests that RPA is not capturing all relevant interaction terms. To address this shortcoming, we resort to functional renormalization group (FRG) [28,67,69] calculations.

### A. Theory

FRG is based on the introduction of an artificial scale into the noninteracting propagator. This artificial scale reduces the interacting system to a solvable system at some initial scale. This initial solvable model can, for example, be the noninteracting case [28] or a DMFT solution [70]. The dependence on the artificial scale is then numerically removed by integrating a hierarchy of flow equations – which is however infinite. In order to treat this numerically, we truncate the hierarchy at the three-particle interactions and additionally drop the self-energy and the frequency dependence of the two-particle vertex. To be able to study the 6-band system without $SU(2)$ symmetry under consideration here, we additionally require using the truncated unity FRG (TUFRG) [43–47] allowing for the treatment of complicated multi-orbital systems [58,71–75]. This level of approximation is also called RPA+-flow, as it includes all different RPA channels and the feedback in-between these (apart from the multiloop class of diagrams [76,77]). The integration has to be stopped once a divergent coupling is encountered, as the assumption that higher vertices are negligible breaks down in this case. To obtain information of the symmetry-broken state, we investigate the leading contributions in the different diagrammatic channels, each encoding different types of orders. The particle-particle (pp) channel encodes superconducting instabilities, the particle-hole (ph) channels (both crossed and direct) encode charge and spin density wave instabilities. Since we will investigate a system with non-negligible SOC, the magnetic and charge channels do not coincide with the crossed and direct particle-hole channel, but have to be obtained by a unitary transformation in spin-space. For a thorough introduction into FRG we refer to Refs. [28,68,69,78]. We use the implementation of the TUFRG as realized in the divERGe library [47] for our calculations. Details of the numerical parameters can be found in Sec. S1C of the Supplemental Material [66]. The inclusion of all diagrammatic channels allows for more realistic interaction parameters (however still reduced from the *ab-initio* values [65,79]) as screening is better captured. These larger interaction energy scales make the results also less prone to small changes of the Fermi surface. For a critical assessment of the applied flavor of FRG we refer to Ref. [58].

### B. Phase diagram

Using the tight-binding models described in Sec. II, we perform FRG calculations to determine the leading electronic instabilities. From these calculations, we can determine not only the energy scale $\Lambda_c^{pp}$ (or $\Lambda_c^{ph}$) of the leading instability, but also its nature. First, we consider the phase diagram for the models without SOC, Fig. 4(a). We find a rich phase diagram with superconducting and magnetically ordered phases. The superconductivity exhibits maxima along a line that connects $\theta \approx 4.5°$ and zero strain $\epsilon$ with zero rotation $\theta$ and $\epsilon \sim 3.75\%$, where the $d_{xy}$ VHs is at the Fermi energy [compare Fig. 1(e)]. For octahedral rotations $\theta > 8°$, no superconductivity is observed, instead we find only magnetic order for the range of parameters considered here. When SOC is introduced, a similar qualitative behavior of the superconductivity is seen [Fig. 4(b)], however now with maxima at smaller values of $\theta$ and $\epsilon$. The most striking difference to the phase diagram without SOC is that magnetic order is practically completely suppressed, and the superconductivity is suppressed in a wider range of the phase diagram. At zero strain, $\epsilon = 0\%$, the superconductivity is suppressed for $\theta > 5.5°$ almost up to $\theta \sim 12°$, and does not recover for experimentally achievable levels of uniaxial strain.

As a function of angle $\theta$, the pairing strength exhibits a maximum for $\theta \approx 3.5°$, before being rapidly suppressed with a minimum between $6°$ and $9°$. Interestingly, the minimum in the pairing interaction persists over a wider range of octahedral rotations than the minimum in the DOS, suggesting that the minimum is not only driven by the SOC induced reduction of the DOS at the Fermi level. This is confirmed by the behavior seen in calculations without SOC, Fig. 4(a). At high octahedral rotation angles $\theta \geqslant 10°$ a small pairing instability is recovered, much smaller than the one found at low rotation angles.

For $\theta = 0°$, we observe an increase of $T_c$ with uniaxial strain $\epsilon$ that is consistent with experiments [30,80].

### C. Superconducting order parameter

An interesting question is how the symmetry of the order parameter is affected by the topological changes of the FS with octahedral rotation. While the RPA has shown a plethora of order parameters, we find only a single leading instability of $d_{x^2-y^2}$ ($B_2$) symmetry (expressed in terms of the one Ru-atom unit cell). This is consistent with results for the bulk from earlier works [36,41,42,58,81,82], further justifying our choice to consider the electronic structure of a free-standing single layer. While the leading order parameter retains its symmetry with increasing rotation, at low rotation angles it is dominant on the $d_{xy}$-orbital, with the largest gap close to the Lifshitz point and with a clear nodal structure. At intermediate angles, where the $d_{xy}$-derived VHs becomes gapped out due to hybridization with the $d_{xz/yz}$ band, superconductivity becomes suppressed and only remerges for significantly larger octahedral rotations but with much smaller pairing strength. At large rotation angles $\theta \geqslant 10°$ we find an inter-orbital $d_{x^2-y^2}$-symmetry with dominant weight on the $d_{xz}/d_{yz}$-derived bands, see Fig. 4(c). The order parameter nominally still retains a sign structure consistent with a $d$-wave symmetry, however it has constant pairing strength on the $d_{xz}$ and $d_{yz}$ bands but with opposite sign and notably without nodes. The main reason for this change is that for a circular Fermi-surface, the $d$-wave component of the particle-particle susceptibility is strongly suppressed at small radii as the $d$-wave form factor is zero at $\Gamma$ and grows proportionally to $\cos(k)$ away from it. Thereby, the $d$-wave order does not benefit from the small pocket near the $\Gamma$ point even though it has a high density of states, as the intra-band pairing susceptibility in the relevant form-factor is essentially zero. This results in the absence of correlation of DOS and observed $T_c$ shown in Fig. 4(d).

## V. EXPERIMENTAL DETECTION OF THE SYMMETRY OF THE SUPERCONDUCTING ORDER PARAMETER

There are only a few experimental techniques that can directly detect the symmetry of the superconducting order parameter. In the cuprate superconductors, early indirect evidence for the symmetry of the order parameter came from measurements of Andreev bound states (for a review see, e.g., [83]), and was supported through the corner-SQUID [84,85] and tricrystal experiments [86], which, however, do not provide direct information about the $k$-space structure of the order parameter, and are hence not universally applicable to resolve the symmetry of the superconducting gap function. The symmetry was later confirmed from neutron scattering [87] and quasiparticle interference (QPI) imaging [11], both techniques that can provide phase sensitive information in scattering ($q$) space. Here, we will concentrate on QPI imaging as a tool to detect not only the magnitude, but also the sign structure of the order parameter. The key signature of the sign structure in QPI is in the phase of the scattering patterns: dependent on the nature of the impurity, magnetic or nonmagnetic, and whether quasiparticles undergo a sign change on scattering, the scattering patterns in real space are either in phase or out of phase between positive and negative bias voltages [11,88,89]. In the following, we will first introduce how the Bogoliubov quasiparticle interference (BQPI) can be modeled, and then the concept of the phase-referenced Fourier transformation for its experimental detection. While in the reconstructed surface layer of Sr$_2$RuO$_4$ superconductivity is completely suppressed, our calculations show that if the octahedral rotation angle can be controlled, superconductivity is expected to re-emerge.

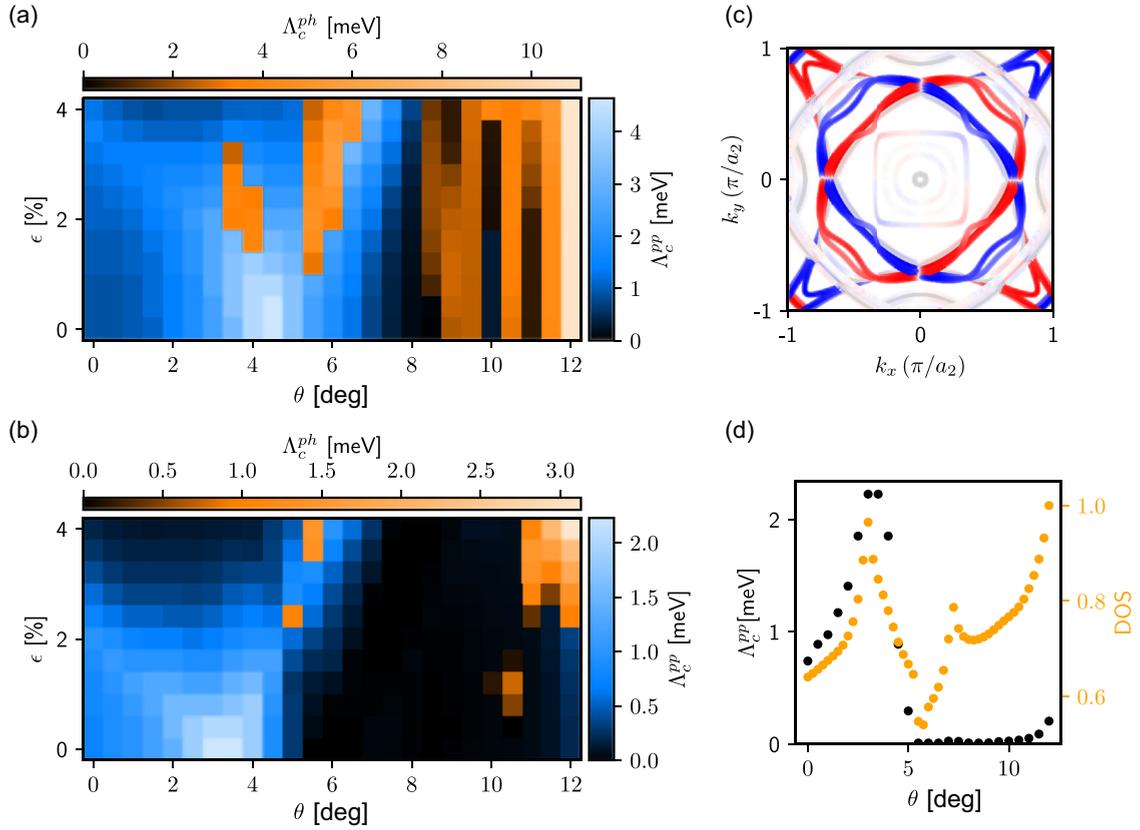

FIG. 4. Phase diagram of Sr$_2$RuO$_4$ as a function of angle $\theta$ and strain $\epsilon$ from FRG. (a) Phase diagram as a function of octahedral rotation $\theta$ and uniaxial strain $\epsilon$ without SOC. The type of the leading phase transition, particle-hole or particle-particle, is encoded in the color, with orange indicating a particle-hole (density-wave) and blue indicating a particle-particle (superconducting) instability. Maxima in the superconducting transition temperature occur along a line connecting $(\theta, \epsilon) = (4.5°, 0\%)$ and $(\theta, \epsilon) = (0°, 3.75\%)$. (b) Corresponding phase diagram with SOC included. The maximum superconducting instability is moved to lower angle $\theta$ and lower values of $\epsilon$. SOC suppresses both instabilities, i.e., the tendency to form magnetic order and the superconducting instability. For all calculations we fix $U = 1.2$eV and $J = 0.14U$. (c) Fermi surface with the superconducting order parameter $\Delta(\mathbf{k})$ for rotation angles $\theta = 0, 2, 4, 6, 8, 10, 12°$. The symmetry remains $d_{x^2-y^2}$ (irrep. $B_2$) for all rotation angles, however for large rotation angles $\theta \geqslant 10°$ becomes quasi $s$-wave, though the pairing strength at large rotation angles is significantly smaller than at low angles (Fermi surfaces for all angles plotted separately in Fig. S4 of the Supplemental Material [66]). (d) Comparison of the critical scale of the divergence of the particle-particle channel (superconducting instability) with the DOS at the Fermi energy, $\rho(E_F)$, showing that the suppression for $\theta > 5°$ is not simply a consequence of the suppressed DOS. For numerical details see Sec. S1C of the Supplemental Material [66].

### A. Calculation of quasiparticle interference

To simulate QPI measurements, we use the continuum Green's function method [90–92] using the St Andrews calcqpi code [93–95]. The quasiparticle interference is obtained using the Green's function of the unperturbed host,

$$\hat{G}_{0,\sigma}(\mathbf{k}, \omega) = \sum_n \frac{\xi_{n\sigma}^\dagger(\mathbf{k})\xi_{n\sigma}(\mathbf{k})}{\omega - E_{n\sigma}(\mathbf{k}) + i\eta}, \quad (2)$$

where $\mathbf{k}, \omega$ are the momentum and energy, and $\xi_{n\sigma}(\mathbf{k})$ and $E_{n\sigma}(\mathbf{k})$ the eigenvectors and eigenvalues of the tight-binding model, and $\eta$ an energy broadening parameter. $\hat{G}_0$ is a matrix in orbital space. From $\hat{G}_0(\mathbf{k}, \omega)$, we obtain the real space Green's function $\hat{G}_0(\mathbf{R}, \mathbf{R}', \omega)$ by Fourier transformation. The Green's function $\hat{G}(\mathbf{R}, \mathbf{R}', \omega)$ of the system including the defect is obtained by using the $\hat{T}$-matrix formalism,

$$\hat{G}_\sigma(\mathbf{R}, \mathbf{R}', \omega)$$
$$= \hat{G}_{0,\sigma}(\mathbf{R} - \mathbf{R}', \omega) + \hat{G}_{0,\sigma}(\mathbf{R}, \omega)\hat{T}_\sigma(\omega)\hat{G}_{0,\sigma}(-\mathbf{R}', \omega),$$
$$(3)$$

where the $\hat{T}_\sigma$-matrix is obtained from

$$\hat{T}_\sigma = \frac{\hat{V}_\sigma}{\mathbb{1} - \hat{V}_\sigma \hat{G}_{0,\sigma}(0,\omega)}, \quad (4)$$

with $\hat{V}_\sigma$ the defect potential. To realistically simulate STM data, the Green's function is required not only on the discrete lattice sites $\mathbf{R}$, but at an arbitrary position $\mathbf{r}$ in space. The Green's function for a continuous spatial coordinate $\mathbf{r}$ is obtained using the continuum transformation

$$\hat{G}_\sigma(\mathbf{r}, \mathbf{r}', \omega) = \sum_{\mathbf{R},\mathbf{R}',\mu,\nu} \hat{G}_\sigma^{\mu,\nu}(\mathbf{R}, \mathbf{R}', \omega) w_{\mathbf{R},\mu}(\mathbf{r}) w_{\mathbf{R}',\nu}(\mathbf{r}'), \quad (5)$$

where the functions $w_{\mathbf{R},\mu}(\mathbf{r})$ describe the Wannier functions corresponding to band $\mu$. These are obtained from the Wannier90 [53] downfolding of the DFT calculations discussed in Sec. II. In particular, they allow realistic modeling of the wave function overlap between the tip and the states in the sample. The QPI is then obtained by calculating the continuum local density of states (cLDOS) from

$$\rho_\sigma(\mathbf{r}, \omega) = -\frac{1}{\pi} \mathrm{Im} \hat{G}_\sigma(\mathbf{r}, \mathbf{r}, \omega). \quad (6)$$

In the following, we assume that the differential conductance $g(\mathbf{r}, V)$ is proportional to the cLDOS, i.e. $g(\mathbf{r}, V) = C\rho(\mathbf{r}, eV)$, with a spatially independent proportionality constant $C$.

Superconductivity is introduced into the real space Hamiltonian using Nambu spinors and introducing the pairing interaction $\hat{\Delta}(\mathbf{R})$,

$$\hat{H}_s = \begin{bmatrix} \hat{H}(\mathbf{R}) & \hat{\Delta}(\mathbf{R}) \\ \hat{\Delta}^\dagger(\mathbf{R}) & -\hat{H}^T(\mathbf{R}) \end{bmatrix}. \quad (7)$$

All calculations shown in this work are for nonmagnetic defects, i.e. $\hat{V} = V_0 \begin{bmatrix} \mathbb{1} & 0 \\ 0 & -\mathbb{1} \end{bmatrix}$ with $V_0 = 1$ eV.

For direct comparison with experimental data, we discuss here the calculated tunneling spectra and spectroscopic maps for the band structure of $Sr_2RuO_4$ using the gap functions obtained from the RPA and FRG calculations (for details see Sec. S4 of the Supplemental Material [66]). In Fig. 5, we show the basic procedure of such a QPI experiment from a calculated conductance map. In quasiparticle interference maps $g(\mathbf{r}, V)$ of the LDOS $\rho_\sigma(\mathbf{r}, eV)$ at tip height $z = 3$ Å of an area with a single defect, Fig. 5(a), one can see the characteristic spatial modulations of the LDOS due to quasiparticle scattering. These encode the information about the electronic structure and the symmetry of the superconducting order parameter. Simulated tunneling spectra [Fig. 5(b)] reveal the superconducting gap, and, if recorded close to the site of the defect, how the superconducting condensate reacts to the presence of the nonmagnetic defect. The Fourier transform of such a QPI map, Fig. 5(c), shows characteristic scattering vectors that can be related to details of the band structure [Fig. 5(d)] and allow extracting the low energy electronic structure [52].

### B. Phase-referenced Fourier transform

Hirschfeld *et al.* have shown that the scattering patterns around nonmagnetic defects in a superconductor provide a robust way to determine the symmetry of the superconducting order parameter in the iron-based superconductors [88]. The method relies on determining the phase of the quasiparticle scattering patterns at different bias voltages, which can be obtained from STM measurements. Determination of this phase from a Fourier transformation of conductance maps requires correcting a global phase factor that is unrelated to the phase shift at the scatterer, but depends on the exact position of the defect within the field of view. This problem can be circumvented by using the phase-referenced Fourier transformation (PR-FFT) [89,96], which allows for the detection of the relative phase between modulations at different energies, and removes the global phase factor. The PR-FFT is calculated from

$$\tilde{g}_{\mathrm{PR}}(\mathbf{q}, eV) = \tilde{g}(\mathbf{q}, eV) \cdot \left( \frac{\tilde{g}(\mathbf{q}, eV_0)}{|\tilde{g}(\mathbf{q}, eV_0)|} \right)^{-1}, \quad (8)$$

where $\tilde{g}(\mathbf{q}, eV)$ is the Fourier transformation of a differential conductance map $g(\mathbf{r}, eV)$ at energy $E = eV$, and $eV_0 = E_0$ is the reference voltage/energy. To study Bogoliubov quasiparticles, it is convenient to set $V_0 = -V$ (where $V = 0$V is the Fermi energy). With this choice, a positive sign of the PR-FFT for a given $\mathbf{q}$-vector indicates that the quasiparticle interference patterns between positive and negative energies are in-phase, whereas a negative sign indicates that they are out of phase. While the normal QPI signal contains information about the momentum space structure of the modulus of the superconducting order parameter, $|\Delta(\mathbf{k})|$, the PR-FFT encodes information about its sign structure, Figs. 5(d) and 5(e). The colors in Fig. 5(e) encode whether a particular scattering vector connects Fermi surface sheets with equal (blue) or opposite (red) sign, and is thus a fingerprint of the sign structure and symmetry of the superconducting order parameter $\Delta(\mathbf{k})$. The PR-FFT has successfully revealed signatures of the symmetry of the order parameter in cuprate [97] and iron-based superconductors [98].

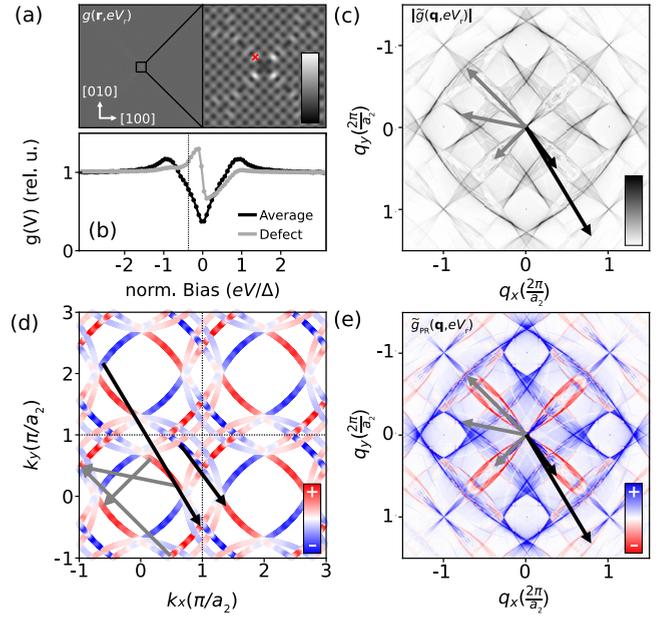

FIG. 5. Quasiparticle interference. (a) Quasiparticle interference map $g(\mathbf{r}, eV)$ calculated in real space from Eq. (6) for $V = -0.46$ mV (ca. $70 \times 70$ nm$^2$). The right half shows a calculation with a higher pixel resolution close to the defect. (b) Spatially averaged continuum LDOS $\langle g(\mathbf{r}, eV) \rangle_\mathbf{r}$ (black) and the cLDOS $g(\mathbf{r}_d, eV)$ at a position close to the defect (red cross in (a), grey). (c) Absolute value of the Fourier transform $|\tilde{g}(\mathbf{q}, eV)|$ of (a). The arrows indicate five scattering vectors. (d) Fermi surface shown over four Brillouin zones. The sign of the gap is encoded in the color. The arrows indicate the scattering vectors shown in (c), where the grey arrows indicate sign-preserving scattering vectors and the black arrows sign-changing scattering vectors. (e) Phase-referenced Fourier transform $\tilde{g}_{\mathrm{PR}}(\mathbf{q}, eV)$ for $V = -0.37 eV/\Delta$, with the arrows as in (b) and (c).

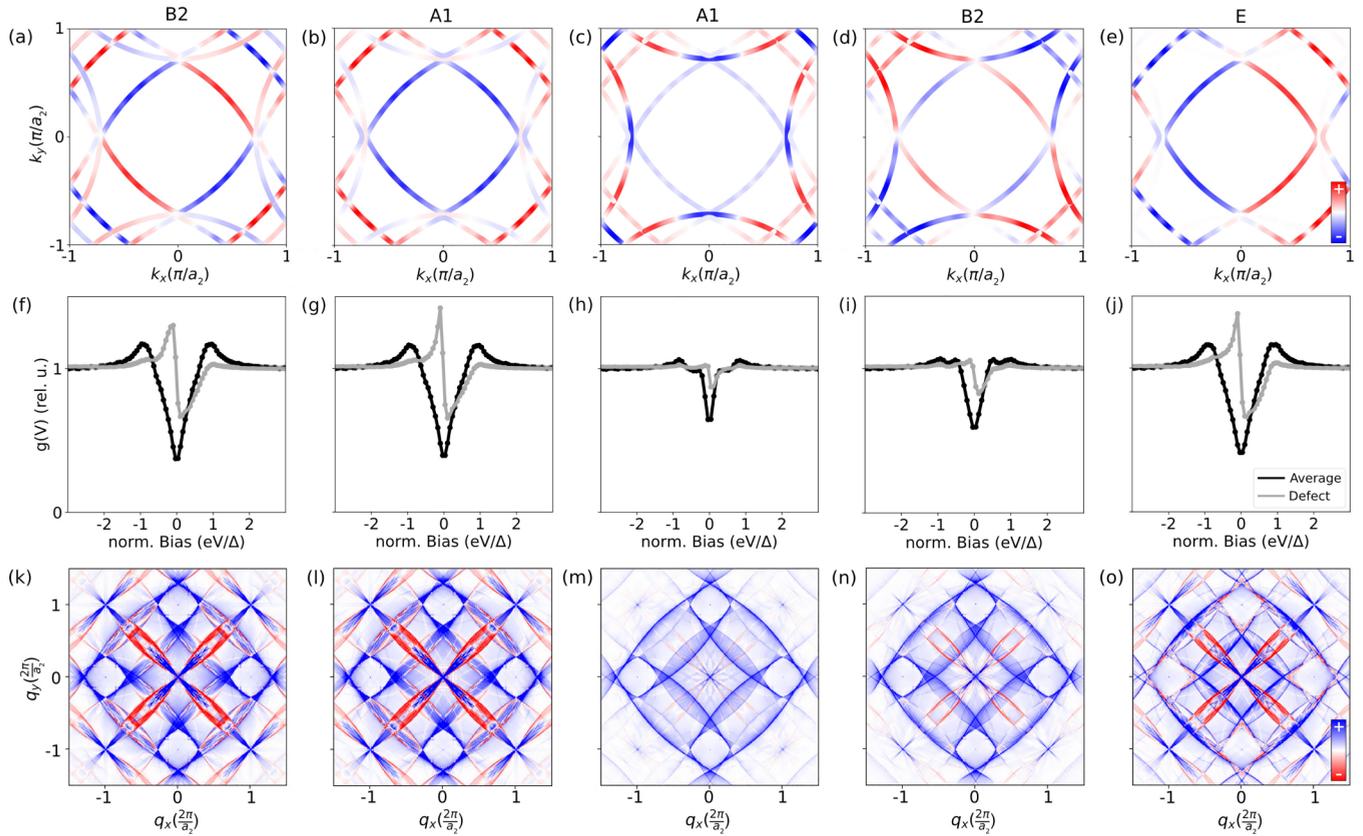

FIG. 6. PR-FFT for the leading five pairing symmetries obtained from the RPA calculation at $\theta = 0°$ without spin-orbit coupling. (a)–(e) Gap functions superimposed on the Fermi surface for gaps with $B_2$ and $A_1$ symmetry dominant on the $d_{xz}/d_{yz}$ bands (a, b), $A_1$ and $B_2$ symmetry dominant on the $d_{xy}$ band (c, d), and an order parameter of $E$ symmetry (e). (f)–(j) Spatially averaged tunneling spectra $\langle g(\mathbf{r}, eV) \rangle_\mathbf{r}$ (black) showing the superconducting gaps (black line) and spectra $g(\mathbf{r}_d, eV)$ near the defect site. All spectra at the defect site show a bound state. (k)–(o) PR-FFT $\tilde{g}_{PR}(\mathbf{q}, eV)$ for $eV = 0.2\Delta$ for the order parameters shown in (a)–(e). Despite the differences in the order parameters, the PR-FFTs for order parameters with $A_1$ and $B_2$ symmetry (a, b and c, d) look very similar for dominant pairing on bands of the same orbital character.

### C. Bogoliubov QPI for different order parameters

#### 1. Results from RPA

In order to understand how different symmetries of the order parameter show up in QPI and in the PR-FFT, it is instructive to initially consider the RPA calculations without SOC, using the identical underpinning normal state band structure without octahedral rotation, $\theta = 0°$, and strain, $\epsilon = 0$. In Figs. 6(a)–6(e), we show the five leading superconducting order parameters obtained from RPA together with simulated tunneling spectra, Figs. 6(f)–6(j), and the PR-FFTs, Figs. 6(k)–6(o). The two leading pairing interactions exhibit dominant pairing on the $d_{xz/yz}$ bands, once with $B_2$ ($d_{x^2-y^2}$) symmetry [Figs. 6(a), 6(f), 6(k)] and once with $A_1$ symmetry [Figs. 6(b), 6(g), 6(f)]. The next two eigenvalues represent dominant pairing on the $d_{xy}$ bands, with either $A_1$, [Figs. 6(c), 6(h), 6(m)] or $B_2$ symmetry. [Figs. 6(d), 6(i), 6(n)]. Finally, Figs. 6(e), 6(j), 6(o) show the results for an order parameter of $E$ ($p$)-symmetry. The signatures of the order parameter, even just in the tunneling spectra, are most obvious when the superconductivity is dominated by the bands of $d_{xz}/d_{yz}$ character [Figs. 6(g), 6(h)], which couple most strongly to the tip of the STM. This is directly reflected in a deeper gap in the tunneling spectrum, compare Figs. 6(f), 6(g) and Figs. 6(h), 6(i). Interestingly, the QPI and the PR-FFTs of the $A_1$ and $B_2$ symmetries are virtually indistinguishable, Figs. 6(k), 6(l) – a consequence of the QPI being dominated by intra-orbital scattering. This remains true for gaps of the same two symmetries on the $d_{xy}$ band. Again, QPI scattering is dominated by intra-band scattering, but the differences in the PR-FFTs remain subtle, Figs. 6(m), 6(n). For a superconducting order parameter with a $p$-symmetry, the PR-FFT [Fig. 6(o)] shows signatures distinct from those of the other order parameters examined here, though notably retaining a $C_4$ symmetry despite the order parameter having only a $C_2$ symmetry. Our results, therefore, show that while the PR-FFT provides valuable information about the sign structure of the order parameter, it cannot always uniquely resolve the symmetry, and deciphering the PR-FFT patterns requires realistic modeling. In Fig. S3 of the Supplemental Material [66], we show QPI calculations for the leading 20 pairing instabilities to cover a larger range of symmetries of the order parameter, including several which exhibit $E$ ($p$) symmetry.

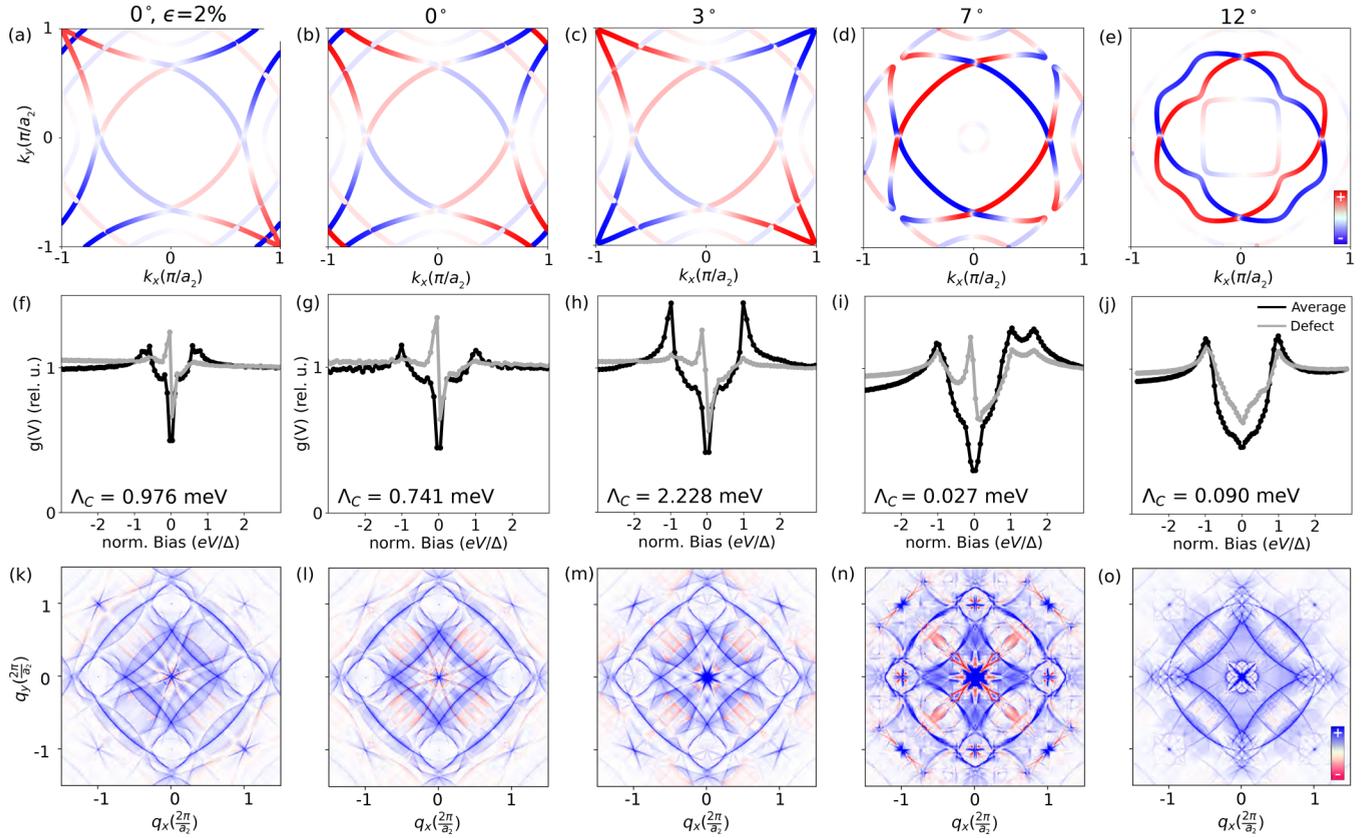

FIG. 7. Gap structure, continuum LDOS and PR-FFTs for the superconducting state found in FRG. (a)–(e) Superconducting order parameter $\Delta(\mathbf{k})$ shown on the Fermi surface for the ($\theta = 0°$, $\epsilon = 2\%$) and ($\theta = 0, 3, 7, 12°$, $\epsilon = 0$) cases, respectively. (f)–(j) Spatially averaged continuum LDOS (cLDOS), $\langle g(\mathbf{r}, eV)\rangle_{\mathbf{r}}$, in black and cLDOS $g(\mathbf{r}_d, eV)$ near the defect site in gray for same (angles $\theta$, strain $\epsilon$) as in (a)–(e). (k)–(o) PR-FFTs $\tilde{g}_{PR}(\mathbf{q}, eV)$ of the cLDOS $g(\mathbf{r}, eV)$ at $eV = 0.2\Delta$.

### 2. Bogoliubov QPI for pairing instabilities from FRG

Having established what differences we might expect in the QPI for different order parameters, we will now consider how the symmetry of the order parameter appears for the superconducting instabilities identified in FRG for the band structure including SOC and as a function of octahedral rotations $\theta$ and strain $\epsilon$. Due to the inclusion of SOC, the quasiparticle interference patterns are notably more complex. As discussed in Sec. IV C, the order parameter retains $d_{x^2-y^2}$ ($B_2$) symmetry [Figs. 7(a)–7(e)] as a function of octahedral rotation, however with a complete suppression of superconductivity when the SOC-induced partial gap encloses the Fermi energy. At the largest rotation angles, where superconductivity reemerges albeit with a significantly smaller pairing interaction than for $\theta < 5°$, the quasi-$s$-wave nature of the order parameter is apparent, with two Fermi surface pockets that exhibit the opposite sign of the order parameter. The simulated tunneling spectra, Figs. 7(f)–7(j), show a two gap structure, with a larger gap forming on the $d_{xy}$ bands, and a smaller gap on the $d_{xz}/d_{yz}$ bands. Upon increasing the octahedral rotation, the gap on the $d_{xy}$-derived band shrinks and vanishes ultimately for large rotation angles. Calculations for more angles are shown in Fig. S4 of Supplemental Material [66]. Even though the symmetry of the gap is retained, the PR-FFT, Figs. 7(k)–7(o), changes significantly at large octahedral rotation angles $\theta \gtrsim 10°$ – due to the absence of sign changes within the $d_{xz}/d_{yz}$ bands and the intra-band scattering being the main driver for the sign structure detected by this method. Not only does the PR-FFT not show any sign-changing features, there is also no bound state in the point $g(\mathbf{r}, V)$-spectra close to the defect. We note that this would change for defects which introduce inter-orbital scattering.

## VI. DISCUSSION

There are three key findings from our study which we will discuss in the following: (1) there is a notable difference in the prediction of superconductivity between RPA and FRG, (2) relatively modest tuning of structural parameters allows increasing the superconducting order parameter by up to a factor of three, and (3) if superconductivity in the surface layer can be recovered, we provide specific predictions for the symmetry of the order parameter and its signatures in QPI.

### A. Differences in order parameter between RPA and FRG

There are notable differences between the predictions of the leading pairing instability obtained from FRG and RPA, exemplified by the very different phase diagrams the two techniques predict, Figs. 3 and 4, which suggest that RPA neglects important contributions to the pairing interaction. This concerns not only the symmetry of the order parameter, but also its stability. RPA predicts that the symmetry of the order parameter changes as a function of octahedral rotation $\theta$, where within FRG, it remains the same throughout the phase diagram. With regards to the pairing strength, RPA suggests an enhancement of the pairing instability even where the DOS at the Fermi energy is reduced due to the SOC-induced partial gap, with the strongest pairing at the largest rotation angles $\theta$ considered here.

In particular, the RPA analysis cannot explain the complete suppression of superconductivity at the surface observed in the experiment. This can be attributed to the intricate interplay of FS nesting, the spin susceptibility and DOS accessible for pairing: The different contributions to the pairing have disparate dependencies on the rotation $\theta$ and strain $\epsilon$, obscuring a direct fingerprint of the DOS at the Fermi level and the energy of the VHs on the critical temperature, since their interplay is not considered. The RPA is highly susceptible to the physics at the Fermi energy due to the small interaction scales required by the rapid divergence of the spin susceptibilities with $U$. Hence, many-body effects can only slightly renormalize the bare susceptibility profile and the influence of the VHs, even though in energetic proximity, is not adequately captured. This decreases both the estimate of the transition temperature, but also the robustness of the $d_{x^2-y^2}$-wave state obtained at small rotation angles in accordance with other work applying RPA to describe superconductivity in Sr$_2$RuO$_4$ [37,39].

The picture is different in FRG. From the noninteracting susceptibility we identify the main driver for the superconducting state to be the nesting between the $d_{xz}/d_{yz}$ bands. Importantly, in the interacting spin-spin susceptibility, the coupling between the 1D bands and the $d_{xy}$ band leads to an amplitude locking: as a result, the extracted pairing interaction is of roughly the same strength for all three bands. The orbital character is therefore determined by the band carrying the largest DOS at the Fermi-level, which is the $d_{xy}$ band. This locking is produced by the renormalization of the bare interaction in the other diagrammatic channels – which is neglected in RPA. This, in combination with the inability of RPA to reproduce the suppression of $T_c$ at the surface, highlights the importance of a diagrammatically unbiased treatment of multiband systems – even if the leading fluctuation is known *a priori*.

### B. Structural control of superconductivity

Maybe the most striking result of our study is the stability and suppression of superconductivity with octahedral rotations, and the demonstration of a new pathway to increase the superconducting transition temperature in this material. We extend the already known behavior with uniaxial strain by adding octahedral rotation as a new dimension. As fixed points that can be compared to experiment, (1) the increase of $T_c$ with uniaxial strain is consistent with experiments [30] and (2) the suppression of superconductivity in the surface layer around octahedral rotation angles of $\theta \sim 8°$ confirms the lack of a superconducting gap in tunneling spectroscopy [13,14,22].

With regards to the latter, for rotation angles larger than $5°$, the calculations reproduce a strong suppression of $T_c$. This is a consequence of the synergistic collaboration of multiple orbitals realizing the superconducting state, which all become partially gapped due to SOC and the octahedral rotation, suppressing the DOS and the pairing interaction. This result also highlights the importance of including SOC for realistic modeling of the ground state – without SOC, the suppression of $T_c$ is much weaker and magnetic order appears as a competing ground state in the phase diagram. Even when including the experimentally observed nematicity [14] of the surface electronic structure, which is mimicked here by strain with $\epsilon = 0.5\%$, the results remain consistent between theory and experiment in the absence of superconductivity.

Our theoretical calculations suggest that the transition temperature can be increased by at least a factor of two at an octahedral rotation, a magic angle, of $3 - 4°$, close to where the Lifshitz transition occurs. Experimental confirmation of this increase would not only provide an important validation of the pairing mechanism, but also bring the superconductivity into a temperature range that is technically more easily accessible for a spectroscopic study of the gap structure, for example for techniques such as photoemission spectroscopy [99].

There are several ways in which octahedral rotations can be introduced in Sr$_2$RuO$_4$: For the surface layer, the octahedral rotation is already present with $\theta \sim 8°$ [18,52,100], and would need to be reduced. This can be achieved, for example, through epitaxial strain in thin films [33], but also through deposition of adlayers, such as alkali metals, which were suggested to reduce the rotation [101]. It is also conceivable that application of biaxial strain in a strain cell can be used, though whether sufficiently large strains can be applied is an open question. Our finding of an increased $T_c$ also provides a possible alternative explanation to uniaxial strain for the 3K phase in Sr$_2$RuO$_4$ crystals with Ru inclusions [102], which will introduce local structural distortions. In the bulk of the material, substitution of Sr by Ca has been shown to result in octahedral rotations, but at the same time suppresses superconductivity already for minute amounts of Ca [34]. For Ca$_{2-x}$Sr$_x$RuO$_4$ with $x = 1.0$, the octahedral rotation is already more than $10°$ [103], and the superconductivity completely suppressed, consistent with the results from FRG. For larger $x$, the rotation cannot be clearly determined anymore [103], however superconductivity remains suppressed until $x = 2$. This apparent contradiction to our results suggests that the inhomogeneity and scattering introduced by Ca is sufficient to suppress superconductivity completely, effectively canceling out the increase we predict here. This is consistent with the well-known sensitivity of superconductivity in Sr$_2$RuO$_4$ to defects [104]. An alternative possibility would be that the phonon mode associated with the octahedral rotation plays a role for the superconductivity, a scenario that is beyond the scope of this study, but which would be an important question for future work.

### C. Symmetry of the order parameter and its experimental detection

Despite the dramatic changes of the Fermi surface, which goes through a Lifshitz transition as a function of rotation angle $\theta$, we find that the superconducting order parameter obtained from FRG is of $d_{x^2-y^2}$ symmetry throughout the phase diagram. Interestingly, while the order parameter has this symmetry all the way up to the highest octahedral rotation angles, beyond 10° it becomes fully gapped, quasi $s$-wave, on the two Fermi surface sheets which dominate the pairing albeit with opposite sign. This transition to a quasi-$s$-wave order parameter is also reflected in the disappearance of impurity bound states in simulated tunneling spectra near nonmagnetic defects.

To enable experimental verification of these predictions for the symmetry of the order parameter, we provide calculations of the BQPI patterns for different rotation angles. These allow phase-sensitive detection of the order parameter in $Sr_2RuO_4$, which, if superconductivity can be stabilized in the surface layer, can guide experiments to identify the symmetry of the superconducting order parameter and resolve this decades-old mystery. The calculations reveal that because of the orbital selection rules of the dominant scattering vectors, not all symmetries of the order parameter can be uniquely distinguished in a QPI experiment. However, in most cases sign changes do leave characteristic signatures in the PR-FFT, which will put significant constraints on the possible symmetries.

The $d_{x^2-y^2}$ symmetry of the order parameter derived from the FRG calculations is consistent with a wide range of experiments. We refer the reader to recent reviews of experimental evidence (e.g., Ref. [105,106]) for different symmetries of the order parameter. We would like to note, however, that discriminating between a sign-changing order parameter with nodes and an anisotropic gap structure with deep gaps, as found both in some unconventional [107] and conventional superconductors [108,109], can be challenging, requiring ultimately truly phase sensitive experiments such as the QPI experiments proposed here.

## VII. CONCLUSIONS

Pinning down the pairing symmetry of $Sr_2RuO_4$ has remained a highly controversial topic, so far with clear and conclusive experimental evidence for a particular symmetry of the order parameter missing [3]. Our results suggest that the superconducting order parameter of $Sr_2RuO_4$ is of $d_{x^2-y^2}$ symmetry, surprisingly stable across the range of structural parameters investigated here. Our calculations reproduce the experimentally found suppression of superconductivity in the surface layer of $Sr_2RuO_4$, which is found to be enhanced by the inclusion of SOC. The absence of said suppression in our RPA calculations suggests that the interplay between different diagrammatic channels is a crucial ingredient to describe this feature. At the same time, we demonstrate a route to increase the superconducting transition temperature $T_c$ by an octahedral rotation of $3-4°$, a 'magic' angle of $Sr_2RuO_4$. We provide specific predictions that can be tested in an experiment for how the symmetry of the order parameter can be detected in BQPI experiments. Beyond the specific case of $Sr_2RuO_4$, this framework which models the phase-referenced Fourier transformation as obtained from Scanning Tunneling Microscopy experiments to determine the sign structure and symmetry of the superconducting order parameter through a combination of FRG and continuum Green's function calculations, is applicable for many unconventional superconductors. The approach pursued here promises a computational route to identify tuning parameters for superconducting properties in quantum materials.

## ACKNOWLEDGMENTS

We thank Andreas Kreisel, Andy Millis, and Roser Valentí for useful discussions. J.B.P. and D.M.K. acknowledge funding by the DFG under RTG 1995, within the Priority Program SPP 2244 "2DMP" — 443273985. J.B.P. acknowledges support by the Deutsche Forschungsgemeinschaft (DFG, German Research Foundation) for funding through TRR 288–422213477 (Project B05). L.C.R., C.A.H., and P.W. gratefully acknowledge support from the EPSRC through EP/R031924/1, L.C.R. and P.W. from the Leverhulme Trust through RPG-2022-315 and P.W. from EPSRC through EP/X015556/1. R.B. was supported from EPSRC through EP/W524505/1. C.A.M. was supported by the Federal Commission for Scholarships for Foreign Students for the Swiss Government Excellence Scholarship (ESKAS No. 2023.0017) for the academic year 2023-24. D.M.K. acknowledges support by the Max Planck-New York City Center for Nonequilibrium Quantum Phenomena. M.D., T.S., and R.T. received funding from the Deutsche Forschungsgemeinschaft (DFG, German Research Foundation) through Project-ID 258499086 - SFB 1170 and through the Würzburg-Dresden Cluster of Excellence on Complexity and Topology in Quantum Matter – *ct.qmat* Project-ID 390858490 - EXC 2147. We acknowledge computational resources provided through the JARA Vergabegremium on the JARA Partition part of the supercomputer JURECA [110] at Forschungszentrum Jülich, as well as HPC resources provided by the Erlangen National High Performance Computing Center (NHR@FAU) of the Friedrich-Alexander-Universität Erlangen-Nürnberg (FAU). NHR funding is provided by federal and Bavarian state authorities. NHR@FAU hardware is partially funded by the DFG – 440719683. This work also used computational resources of the Cirrus UK National Tier-2 HPC Service at EPCC [119] funded by the University of Edinburgh and EPSRC (EP/P020267/1). P.W. gratefully acknowledges the access to and use of computational resources of the HPC cluster Marvin hosted by the Universität Bonn.

J.B.P. performed FRG calculations, L.C.R. created the DFT models and did initial RPA calculations, M.D. did RPA calculations, L.C.R. and P.W. developed code to introduce the superconducting gaps into the Hamiltonians for QPI calculations, R.B. and C.A.M. did QPI calculations, S.C. developed the PR-FFT, PW led and supervised the project. All authors contributed in analyzing the results. J.B.P., L.C.R., M.D., and P.W. wrote the manuscript with input from all authors.